\documentclass[prl,aps,amsfonts,twocolumn]{revtex4}
\def \Dcl {{\Delta_{cl}^{\rightarrow}}}
\def\textbf#1{{\bf #1}}
\def\>{\rangle}
\def\<{\langle}
\def\beq{\begin{equation}}
\def\eeq{\end{equation}}
\def\be{\begin{equation}}
\def\ee{\end{equation}}
\def\ben{\begin{eqnarray}}
\def\een{\end{eqnarray}}
\def\beqa{\begin{eqnarray}}
\def\eeqa{\end{eqnarray}}
\def\eea{\end{array}}
\def\bea{\begin{array}}
\newcommand{\bei}{\begin{itemize}}
\newcommand{\eei}{\end{itemize}}
\newcommand{\bee}{\begin{enumerate}}
\newcommand{\eee}{\end{enumerate}}
\newcommand{\tr}{{\rm tr}}

\def\tr{{\rm Tr}}

\def\>{\rangle}
\def\<{\langle}

\newtheorem{lemma}{Lemma}

\begin{document} 

\title{Classical information deficit and monotonicity on local operations}
\author{Barbara Synak$^{(1)}$, Micha\l{} Horodecki$^{(1)}$}

\affiliation{$^{(1)}$Institute of Theoretical Physics and Astrophysics,
University of Gda\'nsk, Poland}

\begin{abstract}

We investigate classical information deficit: a candidate for measure of classical correlations emerging  from thermodynamical approach
initiated in \cite{OHHH2001}. It is defined as a difference between amount of information that can be 
concentrated by use of LOCC and the information contained in subsystems. We show nonintuitive fact, that one way version of this quantity
can increase 
under local operation, hence it does  not possess property required for a good  measure of classical correlations. 
Recently it was shown by Igor Devetak \cite{devetak},
 that regularised version of this quantity is monotonic under LO. In this context, our result implies  
that  regularization plays a role of "monotoniser".

\end{abstract}
\maketitle

\section{Introduction}
Correlations are a fundamental property of compound quantum distributed system. The 
study of quantum  correlations was initiated by Einstein, Podolsky and Rosen and 
Schr$\ddot{o}$dinger. They were concerned with entanglement - quantum correlation, which are 
nonexistent in classical physics. 
Usefulness of entanglement in Quantum Information Theory to such task 
like teleportation, 
dense coding has motivated extensive study of it. 
However, an exciting subject of characterizing other interesting 
types  
of correlations has emerged. 
 Namely, quantum correlation has been studied  as a notion independent of entanglement \cite{OHHH2001,Zurek02}. 
There were trials to divide  total correlation into classical and quantum part \cite{HendersonVedral,compl},
defined and analysed in \cite{IBMHor2001} and strange properties of classical correlation of quantum states 
were discovered in \cite{Locking}. A measure of classical correlation has also been proposed in \cite{devwin}.

In \cite{OHHH2001} an operational  measure of quantum correlations was proposed. 
It was based on the idea that by using a 
system in  state $\varrho $ one can draw $(N-S(\varrho))kTln2$ of work from a single heat bath,
where N is amount of qubits in state $\varrho $ and $S(\varrho)$ is von Neumann 
entropy of given state. So that information  function given by:
\be
I(\varrho)=N-S(\varrho)
\ee
can be treated as equivalent to work.
 This scenario was used in 
the distributed 
quantum system, where Alice and Bob are allowed to  perform only local 
operations 
and communicate classically with each other (These are so called LOCC operations) to concentrate 
information 
contained in the state to local subsystems. 
For nonclassical states the amount of work drawn by LOCC (equivalently amount of information $I_{LOCC}$
we can concentrate  by LOCC operation into local subsystem)   is usually 
 smaller than work  exactable by global operations (equivalently information $I_{GO}$, to which we have access by using  global 
operation). 
The resulting difference
$\Delta=I_{GO}-I_{LOCC}$ called information deficit or work deficit and it
accounts for 
the part of correlation that must be lost during 
classical communication, thus describe purely quantum  correlation.
Similar definitions we can apply for one-way (Alice to Bob) information deficit 
$\Delta^\rightarrow$. It differs from $\Delta $ by using in definition one way {LOCC} operation instead 
(two way) LOCC. (For  one way, from Alice to Bob, LOCC operations only communication from Alice to Bob is allowed.)

In \cite{compl} a complementary quantity, that could account for classical correlation was defined - 
classical information deficit $\Dcl$. 
\be
\Delta_{cl}=I_{LOCC}-I_{LO}
\ee
where $I_{LO}$ is amount of information accessible by using only local operations performed 
on $N_A$ qubits of subsystem A and $N_B$ qubits of subsystem B. 
(i.e. $I_{LO}=N_A-S(\varrho_A)+N_A-S(\varrho_B)$.
One can see, that the two measures of correlations 
add up to quantum mutual information given by:
\be
I_M=S(\varrho_A)+S(\varrho_B)-S(\varrho_{AB})
\ee
where $S(\varrho)$ is the von Neumann entropy of  state $\varrho$ and 
 $\varrho_{A(B)}=\tr_{B(A)} \varrho_{AB}$, i.e. we have:
\be
\Delta_{cl} + \Delta =I_M 
\ee

Analogously, we have the 
 following formula for one-way version of classical information deficit:
 \be
\Dcl(\varrho_{AB})= I_{LOCC}- I^\rightarrow_{LO}
\ee

In this paper we show a nonintuitive fact, i.e. that one-way classical information deficit 
$\Dcl(\varrho_{AB})$ is 
not a measure of classical correlation, because can {\it increase} under local 
operation. 
Remarkable, it was recently shown by Devetak \cite{devetakC} that quantity,
which is equal to regularised classical information deficit $\Dcl^\infty (\varrho_{AB})$ 
 is monotonic under LO. Combining those results with ours, we obtain, that regularisation may play a role of monotoniser.
 An  asymptotic version of a function may be monotonic, even though one copy version is not.

\section{Formula for $\Dcl$ and comparison with Henderson-Vedral measure}

In this section we provide formula for  $\Dcl$ and compare
it with measure of classical correlation introduced by Henderson and Vedral. 
To this end we have to  determine formula for the maximal amount of information, 
which can be concentrated to subsystems via a protocol,
in which one-way classical communication is allowed. 
The most general such protocol is the following.  
Alice makes a measurement on her part of state and tells her results to 
Bob. 
The amount of concentrable  information is then equal to the  information of 
Alice 
plus average final information of Bob.
The protocol transforms the state in following way:
\be
\varrho_{AB} \to \varrho_{AB}'=\sum_i P_i \otimes I \varrho_{AB} 
P_i \otimes I 
\ee
where   $p_i$ given by
\be
p_i=\tr ( P_i \otimes I \varrho_{AB} P_i \otimes I )
\ee
is probability that
Bob gets  the state $\varrho_i^B$, which is of  the form:
\be
\varrho_i^B=\tr_A (P_i \otimes I \varrho_{AB} P_i \otimes I )/p_i
\ee
 and $\{P_i\}$ are projectors constituting von Neumann measurement.
Usually, in LOCC paradigm one would allow for  POVM. However POVM requires adding ancillas, which 
we have to take into account, if we are estimating the amount of information that we can 
concentrate. Thus, we include from very beginning all needed  ancillas and 
consider von Neumann measurement. In such a way we take into account  POVM's, too. (There is an open question, if it pays to 
add ancillas at all, we will discuss this later.)

The amount of information $I(\mathcal{P})$, 
which can be concentrated into subsystems in one-way protocol $\mathcal{P}$  is thus equal to:
\ben
I(\mathcal{P}) &=& I_A^{out}+\overline{I_B^{out}}\\
&=& N_A-S(\varrho_{A}') + N_B-\sum_i p_i S(\varrho_{B}^i) \\
&=& N-S(\varrho_{A}') -\sum_i p_i S(\varrho_{B}^i)
\een
where $N_{A(B)}$ is amount of qubits in Alice (Bob) part of state, $(N_A + N_B = N)$,
 $\varrho_{A}'=\tr_B \varrho_{AB}'$. The maximal information that can be concentrated
 by one-way protocols $\mathcal{P}^\rightarrow $ is denoted by $I^\rightarrow$:
\ben
I^\rightarrow=\sup_{\mathcal{P}^\rightarrow } I(\mathcal{P}^\rightarrow )
\een 
Having formula for $I^\rightarrow$ we can express $\Delta^\rightarrow_q$ as:
\ben \nonumber
\label{delta}
\Delta^\rightarrow_q &=& N-S(\varrho_{AB}) -\sup_{\{P_i\}} \{(N-\sum_i p_i S(\varrho_B^i)-S(\varrho'_A)\}\\ 
&=& \inf_{\{P_i\}}\{ \sum_i p_i 
S(\varrho_B^i)+S(\varrho'_A)\}-S(\varrho_{AB})
\een
Since $\Dcl$ is equal to the difference between total information $N-S(\varrho_{AB})$ and $I^\rightarrow$, we obtain:
\ben \nonumber
\label{one-wayclass}
\Dcl(\varrho_{AB}) 
= \sup_{\{P_i\}} [\{S(\varrho_A) - S(\varrho'_A)\} \\ 
+ \{S(\varrho_B) - \sum_i p_i S(\varrho_B^i)\}] 
\een
where the supremum is taken over all local dephasings on Alice's side.
Note that protocol is determined by choosing Alice's measurement.
Note also, 
that the optimal measurement  is a complete one, i.e. $P_i$ can be chosen to be one dimensional projectors.
 Indeed, given any 
incomplete measurement, Alice can always  refine it, in  such a way, that her entropy will not increase,
and of course, any refinement
will not increase 
Bob's average entropy.
In eq. (\ref{one-wayclass}), we have distinguished two terms. The second term

\[ 
S(\varrho_B) - \sum_i p_i S(\varrho_B^i)
\]
shows the decrease of Bob's entropy after Alice's measurement. 
The first one 
\[
 S(\varrho_A) - S(\varrho^{'}_A)
\]
denotes the cost of this process on Alice side,
and is non-positive. It vanishes only if Alice measures 
in the eigenbasis of her local density matrix \(\varrho_A\). Thus,
the expression for \(\Dcl\) is very similar to the measure of 
classical correlation introduced by Henderson and Vedral \cite{HendersonVedral}: 
\be
\label{HendersenVedral}
C_{HV}(\varrho_{AB})=\sup_{P_i}( S(\varrho_B) - \sum_i p_i S(\varrho_B^i) ).
\ee
Originally, in  definition of $C_{HV}$ the supremum was taken over POVMs, 
but as mentioned, we 
take the state acting already on a suitably larger Hilbert space, unless stated otherwise explicitly.
The difference between the Henderson-Vedral classical correlation 
measure 
and the one given in eq. (\ref{one-wayclass}) is that the former does 
not include Alice's entropic cost of performing dephasing. Hence in general
\[
\Delta_{cl}^{\rightarrow} \leq C_{HV}
\]

\section{When $\Delta_{cl}^{\rightarrow}$ can be equal to $C_{HV}.$}

In this section we prove the following lemma:

\begin{lemma}
\label{deltaHV_and_deltaone-waycc}

Let $\varrho_{AB}$ be any bipartite state. Then 
$C_{HV}(\varrho_{AB})=\Delta_{cl}^{\to}(\varrho_{AB}) $ if and only if there exist projectors $\{P_i\}$ such that they commute with 
$\varrho_{A}(= tr_{B} \varrho_{AB})$ and  they are optimal for both 
 $C_{HV}$ and $\Delta_{cl}^{\to}$ for the state 
$\varrho_{AB}$ .
\end{lemma}
\textbf{Remark} 1 Note, that eigenbasis  of $\varrho_A$ may not be unique.

\textbf{Proof.}
For specific measurement , let us use the following notation:
\ben
\label{smalldeltaHV}
c_{HV}&=&S(\varrho_{B})-\sum_{i}p_{i}S(\varrho^B_{i})\\
\label{smalldeltaone-waycc}
\delta_{cl}^{\to}&=&S(\varrho_{A})-S(\varrho '_{A})+S(\varrho_{B})- 
\sum_{i}p_{i}S(\varrho ^B_{i})
\een
The quantity $c_{HV}$ and $\delta_{cl}^{\to}$ are functions of state and a measurement. We have
\ben
C_{HV}  &=& \sup_{\{P_{i}\}}c_{HV} \\
\Delta_{cl}^{\to} &=& \sup_{\{P_{i}\}}\delta_{cl}^{\to}.
\een

$"\Rightarrow"$ Let us proof the ``only if'' part:
Suppose that 
\[
C_{HV} = \Delta_{cl}^{\rightarrow}
\]
Consider measurement (i)  which  achieves $C_{HV}$ and measurement  
(ii), which
achieves $\Delta_{CL}^{\to}$. 
Let $c_{HV}^{i(ii)}$ be the values $c_{HV}$ for measurement $i(ii)$ and $S(\varrho'^{(ii)}_A)$ 
is Alice's part  entropy after measurement $(ii)$. Then
\be
\label{r}
\Delta_{cl}^{\to} = S(\varrho_{A})-S(\varrho'^{(ii)}_{A})+c_{HV}^{(ii)}=c_{HV}^{(i)}=C_{HV}\ee
We know that for arbitrary measurement $S(\varrho_{A})-S(\varrho '_{A})\leq 0$ \cite{W78} and 
$c_{HV}^{(ii)}\leq c_{HV}^{(i)}$.
If we want the equality (\ref{r}) to hold, then  it must be that 
\be
\label{ent}
S(\varrho_{A})-S(\varrho_{A}^{(ii)})= 0 \quad \textrm{and} \quad c_{HV}^{(ii)}=c_{HV}^{(i)}
\ee
It follows that measurement $(ii)$ is also optimal for $C_{HV}$. Moreover, notice, that this measurement is made in eigenbasis,
 otherwise it would increase entropy $S(\varrho '_{A})$ violating eq. (\ref{r}).

$"\Leftarrow "$ 
The "if" proof is obvious. Since we assume that the measurement achieving $C_{HV}$ and $\Delta_{CL}^{\to}$ is the same 
and is made in eigenbasis of $\varrho_A$, so then $S(\varrho_{A})-S(\varrho'_{A})=0$, so that  $\Delta_{cl}^{\to}$  and $C_{HV}$
must be equal.  
This ends the proof of lemma.

\section{When $\Delta_{cl} ^{\to}$ can increase under local operation.}

 \begin{lemma}
 \label{l2}
 If $ \Delta_{cl}^{\to}\neq C_{HV}$
then the quantity $\Delta_{cl}^{\to}$ can be increased
by local operations.
\end{lemma}

Therefore let us  assume that $\Delta_{cl}^{\to} < C_{HV}$ for the state $\varrho_{AB}$.
(Recall that, in general, $\Delta_{cl}^{\to} \le C_{HV}$.)
Let us consider an optimal  measurement $\{P_{i}^{HV}\}$ achieving $C_{HV}$.
After the measurement, the  state is of the form
\[
\varrho'_{AB}=\sum_{i}p_{i} P_{i}^{HV} \otimes \varrho_{i}^{B}.
\]
We know that $C_{HV}$ cannot increase after local operations \cite{HendersonVedral}.
Then 
\be
C_{HV}(\varrho_{AB}) \leq C_{HV}(\varrho'_{AB}) = S(\varrho_B) - 
\sum_i p_i 
S(\varrho_i^B)
\ee
so that  $\{P_{i}^{HV}\}$ is an optimal measurement  for the 
state 
$\varrho '_{AB}$ also.
{\it Now if we repeat the measurement 
$\{P_{i}^{HV}\}$ on 
$\varrho_{AB}'$
we get the same value of $C_{HV}(\varrho_{AB}')$ as before since 
$\varrho_{AB}'$ and the created Bob ensemble do not change
under that particular measurement.}
Thus
\[
C_{HV}(\varrho_{AB}) = C_{HV}(\varrho '_{AB}).
\]
Note that $\{P_{i}^{HV}\}$ corresponds to the eigenbasis of  
$\varrho '_{A}$,
where $\varrho '_{A}$ is the reduced matrix of $\varrho'_{AB}$.
Then
\[
\Delta_{cl}^{\to}(\varrho '_{AB}) = C_{HV}(\varrho '_{AB})
\]
so that
\[
\Delta_{cl}^{\to}(\varrho '_{AB}) > \Delta_{cl}^{\to}(\varrho_{AB})
\]
i.e. $\Dcl$ increase after local operation of dephasing by $P_i$.

Having Lemma \ref{l2} the question is whether there exist states for 
which$\Delta_{cl}^{\to} \neq C_{HV}$. We know that in such case, 
for such states there should not  
exist any measurement optimizing both $\delta_{cl}^{\to}$ and $c_{HV}$, 
which is made in eigenbasis. Equivalently, there should not exist 
a measurement that  optimises $C_{HV}$, which is made in eigenbasis of $\varrho_A$.
To show this, the following results  of Schumacher and 
Westmoreland \cite{SW99} and
King, Nathanson and Ruskai \cite{KingNR2001}
connected with  classical capacity of a quantum channel are helpful.

Suppose a source produces  states \(\varrho_k\) with probabilities \(p_k\).
 For this ensemble, the authors in Ref. \cite{SW99, KingNR2001}
 considered a quantity called Holevo quantity, defined as 
\[
\chi = S(\varrho) - \sum_{k} p_{k} S(\varrho_{k})
\]
where 
\[
\varrho  =  \sum_{k} p_{k} \varrho_{k}.
\]
They were interested in maximizing $\chi$ for the output ensemble
$\{p_{k}, \Lambda (|\psi_{k}\rangle \langle \psi_{k}|)\}$, 
where $\Lambda$ is  fixed completely positive map (channel).

It turns out  that for some channels, to maximize $\chi$, one needs 
a non-orthogonal input  ensemble. This was first shown by Fuchs 
\cite{Fuchs97}.
An example of such a channel  is given by the following map \cite{SW99}:
\ben
\label{eq-kanal1}
\Lambda_{1}(\varrho) &=& A_{1}\varrho A_{1}^{\dagger}+A_{2}\varrho 
A_{2}^{\dagger}\\ \nonumber
\textrm{where} \quad A_{1}&=&\sqrt{\frac{1}{2}}|1\>\<1|+|0\>\<0|\\ 
\nonumber
A_{2}&=&\sqrt{\frac{1}{2}}|0\>\<1| \nonumber
\een
where $\left\{|0 \rangle, |1 \rangle \right\}$ is the standard basis in $\mathbb{C}^{2}$.                
For this channel,  maximum $\chi$ is obtained for non-orthogonal input states.

On the other hand, it has been recently shown \cite{KingNR2001}  that 
sometimes the number of states in the optimal 
ensemble must be 
greater than dimension of the system. 
An example is the map given by
\be
\label{eq-kanal2}
\Lambda_{2}(\varrho) = \frac{1}{2} \Big(I +
 [0.6 \mathnormal{w}_{1}, 0.6 \mathnormal{w}_{2}, 0.5 + 0.5 
\mathnormal{w}_{3} ]. 
\vec{\sigma} \Big)
\ee
where
\[ 
\varrho = \frac{1}{2} (I + \vec{\mathnormal{w}} \vec{\sigma})
\]
and $\vec{\mathnormal{w}}$ = $(\mathnormal{w}_{1}, \mathnormal{w} 
_{2},\mathnormal{w}_{3})$ with
$\vec{\sigma} = (\sigma_{1}, \sigma_{2},\sigma_{3})$ and $\sigma_{i}$ 
being 
the  Pauli matrices.
In this case $\chi$ is maximized by a three component ensemble.

The above examples can lead us to 
a bipartite state \(\varrho_{AB}\), for which \(C_{HV}\) is not  
achieved by the measurement in the eigenbasis of \(\varrho_A\). (Note 
that 
these examples act only as indications.
The results of channel capacities are not  used directly, although such 
a 
direct connection is not ruled out.)

More precisely, given any channel and ensemble, we will construct some
bipartite state
and a measurement on one of its subsystems. We will expect
that the measurement will give high value of  $c_{HV}$ on that state.
In  particular, if  the ensemble is two component but non-orthogonal we obtain  a better
value of $c_{HV}$ than the eigenbasis measurement then  
the optimal measurement for  
\(C_{HV}\) on Alice's part of the state will not be in the eigenbasis 
of 
\(\varrho_A\).
Moreover, if the  ensemble is three component and the measurement will give a better 
value than any von Neumann measurement, the optimal measurement
   for attaining $C_{HV}$
will not be a von Neumann measurement, but POVM. We show that it is indeed the case in both situations.

Let us now present our  construction of the state and measurement 
from a given channel $\Lambda $ and ensemble $\{p_i,\psi_i\}$. We will first exhibit two ways of obtaining 
ensemble $\{p_i,\varrho_i\}$ from a pure bipartite state $\psi_{AB}$, where $\varrho_i=\Lambda(\psi_i) $.
Let $\psi_{AB}$ be a state for which 
\[
\tr_{A} |\psi_{AB} \rangle \langle \psi_{AB}| 
= \sum_{i} p_{i} |\psi_{i}\rangle \langle\psi_{i}|.
\]
One can write it in the form 
$|\psi_{AB} \rangle = \sum_{i} \sqrt{p_i} | i \rangle |\psi_{i}\rangle $, 
where $|i\>$ are orthogonal.
Note that when we make a measurement in the basis $| i \rangle$ at 
Alice's 
side,
the ensemble $\{ p_{i}, \psi_{i} \}$ is created at Bob's side. 
Then one obtains ensemble $\{p_i,\varrho_i\}$ by letting $\psi_i$ to evolve through  the channel  $\Lambda $.
But one can achieve $\{ p_{i}, \varrho_{i} \}$ in a different way.
First, the state $\psi_{AB}$ is prepared and the operation $I_{A}\otimes \Lambda_{B}$ is performed, producing state $\varrho_{AB} $:
\[
\varrho_{AB}=(I_{A}\otimes \Lambda_{B})(	\psi_{AB} )
\]
Then  Alice makes the  measurement 
in the basis \(\left|i\right\rangle\) and this produces  the
ensemble $\{ p_{i}, \varrho_{i} \}$ at Bob's site.
The connection between the scenarios is illustrated by the commuting diagram 
below.
Starting from $\psi_{AB}$, we can achieve the ensemble $\{ p_{i}, \varrho_{i} \}$
in two ways.
\vspace{10 mm}

\be
\begin{array}{rccc}
\psi_{AB}&\mathop{\vector(1,0){70}}\limits^{I_{A}\otimes 
\Lambda_{B}}&\varrho_{AB} \\
\line(0,-1){30}&&\line(0,-1){30}\\  \mbox{\emph{M}}_{A}&&\mbox{\emph{M}}_{A} \\
\vector(0,-1){30}&&\vector(0,-1){30}\\
\{ p_{i}, \psi_{i} \}_{B}&\mathop{\vector(1,0){70}}\limits^{\Lambda}&\{ p_{i}, \varrho_{i} \}_{B}
\end{array}
\ee

\vspace{10 mm}
Here $M_{A}$ denotes the measurement by Alice and $\{ * , * \}_{B}$ 
denotes the corresponding ensemble at Bob's site. 
If we want to find the needed state $\varrho_{AB}$, for which $\Dcl\not= C_{HV}$, we should construct pure state 
$\psi_{AB}$ 
and then perform operation $I_{A}\otimes \Lambda_{B}$.
First we use the channel (given by eq. (\ref{eq-kanal1})) and ensemble 
from 
Ref.
\cite{SW99} to obtain $\varrho_{AB}$ 
for which  $c_{HV}$ for some measurement is greater than for measurement in  
eigenbasis.

An example of a non-orthogonal ensemble,
for the channel \(\Lambda_1\) given by eq. (\ref{eq-kanal1}),
 which gives greater $\chi$,  than any 
orthogonal one is
$\{\{\frac{1}{2},\psi_{0}\}, \{\frac{1}{2},\psi_{1}\}\}$,
where
\begin{eqnarray}
|\psi_{0}\rangle &=& \frac{1}{\sqrt{2}}(|0 \rangle + | 1 \rangle)\\
|\psi_{1}\rangle &=& \frac{4}{5} |0 \rangle + \frac{3}{5} |1 \rangle.
\end{eqnarray}
Then we have  
\begin{equation}
\varrho_{AB} =(I^{A}\otimes \Lambda^B_{1}) |\psi_{AB} \rangle \langle \psi_{AB}|
\label{st}
\end{equation}
where 
\[
\left|
\psi_{AB}\right\rangle = 
\frac{1}{\sqrt{2}}(\left|0\right\rangle_A\left|\psi_0\right\rangle_B + 
\left|1\right\rangle_A\left|\psi_1\right\rangle_B)
\]
                                                                                                                                                      
Now, we can check directly that for Alice's measurement in basis $| 0 \rangle$,$| 1 \rangle$ 
(which prepares the non-orthogonal ensemble 
$\{\{\frac{1}{2},\psi_{0}\}, \{\frac{1}{2},\psi_{1}\}\}$ on Bob's side), the 
Henderson-Vedral quantity \(c_{HV}\) (see eq. (\ref{smalldeltaHV})) 
attains the value
$c_{HV}^{(1)}=0.45667$. But for Alice's measurement in the eigenbasis of
 \(\varrho_A\), \(c_{HV}\) attains the value $c_{HV}^{(2)} = 0.3356$.
Therefore  $c_{HV}^{(1)} > c_{HV}^{(2)}$, i.e. there exists Alice's measurement 
which gives better value for \(c_{HV}\) than the measurement in the 
eigenbasis of \(\varrho_A\).
The optimal measurement is therefore clearly not in eigenbasis. 
{\it This fact, as follows from Lemma 1,  implies that
for state given by formula (\ref{st})
one has  $C_{HV} \neq \Dcl$
and, even more remarkably, as follows from Lemma \ref{l2}, 
for this state $ \Dcl$ {\it increases} under local measurements.}
The operation that increase $ \Dcl$  is Alice's dephasing in basis
 $|0\>,|1\>$.

We now  use the  results of Ref. \cite{KingNR2001} to find next example of state for 
which $C_{HV}\neq \Dcl $.                     
The three component ensemble for which $\chi$, for the channel 
\(\Lambda_2\)
given by eq. (\ref{eq-kanal2}), is greater  than that for any two 
component 
ensemble is
$\{p_{i}, | \phi_{i}\rangle\}$
where $p_{0}$ = 0.4023, $p_{1}$ = $p_{2}$ = 0.29885 and 
\[
\begin{array}{rcl}
|\phi_{0}\rangle &=&| 0 \rangle\\
|\phi_{1}\rangle &=& a| 0 \rangle + b| 1 \rangle\\
|\phi_{2}\rangle &=& a| 0 \rangle - b| 1 \rangle
\end{array}
\]
with a = 0.0701579, b = 0.821535.
Then by our prescription,
{\it the state for which  POVM is better than 
any von Neumann measurement, as far as $C_{HV}$ is concerned, is}
\[
\varrho_{AB} =(I^{A}\otimes \Lambda^B_{2}) |\phi_{AB} \rangle \langle \phi_{AB}|
\]
where
\[
|\phi_{AB} \rangle = \sum_{i=0}^{2} \sqrt{p}_{i}|i \rangle |\phi_{i} \rangle.
\]

Again, as from Lemma 1 and Lemma 2, for this state $ \Dcl \neq C_{HV}$, hence $\Dcl$ can be increased by Alice's
 dephasing in basis  $\{| 0 \rangle, | 1 \rangle, | 2 \rangle\}$, which can be treated a a POVM,
 since Alice's subsystem has rank two, so it is  efficiently qubit. For the  measurement in the basis $|i \rangle$ (when the ensemble 
$\{p_{i}, | 
\phi_{i}\rangle\}$
is prepared on Bob's side),  $c_{HV}$ attains the value 
$\delta^1_{HV}$ = 0.32499.
For von Neumann measurements, $c_{HV}\le 0.321915$.
Equality is obtained for measurement in eigenbasis.

{\it Finally, let us show that POVMs that are good for 
$C_{HV}$ can be very bad for
$\Delta_{cl}$}. 
One can check that the same POVM which gives high value for $c_{HV}$, gives $\delta_{cl}^{\rightarrow} < 0 $.
Therefore a POVM which is good for $c_{HV}$ can be
very bad for $\delta_{cl}^{\rightarrow} $.

We have checked that for $\delta_{cl}^{\rightarrow}$, the best von 
Neumann 
measurement is in eigenbasis. 
Then \(\delta_{cl}^{\rightarrow}\) attains the value 
$\delta_{cl}^{\rightarrow (vN)}\approx  0.321915.$
 
This example indicates  that $\Delta_{cl}^{\rightarrow}$ might be   such a quantity for 
which POVMs are not helpful. We conjecture that it can be truth.

 \section{Discussion} 
 In this paper we have considered classical information deficit $\Delta_{cl}^{\rightarrow}$ defined as
 difference between amount of information that can be concentrated by LOCC and information concentrable by LO.
 It  is equal to
 difference  between measure of total correlation and measure of quantum correlation present in state.
It was  reasonable to expect  that it should be a  measure of classical 
correlation.
 We have shown, that it is not true, because $\Delta_{cl}^{\rightarrow}$ can increase under local operation.
 We have proved it through comparison it with measure of classical correlation by
Henderson-Vedral.
 We  based on lemma, which tells us, when these quantities can be 
equal. 
We showed that, if they are different, then 
 $\Delta_{cl}^{\rightarrow}$ can increase under local actions.
 The last thing we did is finding examples of states for which 
$\Delta_{cl}^{\rightarrow}\neq C_{HV}$.
 We also exhibited example, where POVM is very good for  $C_{HV}$, but completely bad for $\Delta_{cl}^{\rightarrow}$.
This suggest that POVM's may be not helpful in one-way protocol of localizing information.
 This would be compatible with result for two-way
protocols, were borrowing ancillas does not help in concentrating information \cite{huge}. 
The above results would mean that $\Dcl$  is useless as far as classical correlation of quantum states 
are concerned. Fortunately,  it is not the case.

Recently it was shown in \cite{devetak} that regularized version 
of 
$\Delta_{cl}^{\rightarrow}$ is a measure of classical correlation, because it is
equal to distillable common randomness \cite{devetakC}, which in fact is equal to regularised $C_{HV}$.
 Since the latter is monotonic under local operation, then $\Dcl$
 if regularised is monotonic, too.
 It is very 
puzzling fact, 
that we have quantity, which defined  
for one copy of state can increase after local operations, but its regularized version not.
Thus, according to our results, the regularization plays a role of "monotoniser" in this case.


{\bf Acknowledgments}:
We thanks Ryszard Horodecki for helpful discussion and Aditi Sen(De) for help 
in making improvement some part of this letter.
This work is supported by EU grants RESQ, Contract No. IST-2001-37559 
and  
QUPRODIS, Contract No. IST-2001-38877 and the Polish Ministry of Scientific
 Research and Information Technology under the (solicited) grant No
 PBZ-MIN-008/ P03/ 2003.

\bibliography{refbasia}

\begin{thebibliography}{14}
\expandafter\ifx\csname natexlab\endcsname\relax\def\natexlab#1{#1}\fi
\expandafter\ifx\csname bibnamefont\endcsname\relax
  \def\bibnamefont#1{#1}\fi
\expandafter\ifx\csname bibfnamefont\endcsname\relax
  \def\bibfnamefont#1{#1}\fi
\expandafter\ifx\csname citenamefont\endcsname\relax
  \def\citenamefont#1{#1}\fi
\expandafter\ifx\csname url\endcsname\relax
  \def\url#1{\texttt{#1}}\fi
\expandafter\ifx\csname urlprefix\endcsname\relax\def\urlprefix{URL }\fi
\providecommand{\bibinfo}[2]{#2}
\providecommand{\eprint}[2][]{\url{#2}}

\bibitem[{\citenamefont{Oppenheim
  et~al.}(2002{\natexlab{a}})\citenamefont{Oppenheim, Horodecki, Horodecki, and
  Horodecki}}]{OHHH2001}
\bibinfo{author}{\bibfnamefont{J.}~\bibnamefont{Oppenheim}},
  \bibinfo{author}{\bibfnamefont{M.}~\bibnamefont{Horodecki}},
  \bibinfo{author}{\bibfnamefont{P.}~\bibnamefont{Horodecki}},
  \bibnamefont{and}
  \bibinfo{author}{\bibfnamefont{R.}~\bibnamefont{Horodecki}},
  \bibinfo{journal}{Phys. Rev. Lett} \textbf{\bibinfo{volume}{89}},
  \bibinfo{pages}{180402} (\bibinfo{year}{2002}{\natexlab{a}}),
  \eprint{quant-ph/0112074}.

\bibitem[{dev()}]{devetak}
\bibinfo{note}{I.Devetak. In preparation}.

\bibitem[{\citenamefont{\.Zurek}(2003)}]{Zurek02}
\bibinfo{author}{\bibfnamefont{W.~H.} \bibnamefont{\.Zurek}},
  \bibinfo{journal}{Phys. Rev. Lett. A} \textbf{\bibinfo{volume}{67}},
  \bibinfo{pages}{012320} (\bibinfo{year}{2003}), \eprint{quant-ph/0202123}.

\bibitem[{\citenamefont{Henderson and Vedral}(2001)}]{HendersonVedral}
\bibinfo{author}{\bibfnamefont{L.}~\bibnamefont{Henderson}} \bibnamefont{and}
  \bibinfo{author}{\bibfnamefont{V.}~\bibnamefont{Vedral}},
  \bibinfo{journal}{Jour. of Phys. A} \textbf{\bibinfo{volume}{34}},
  \bibinfo{pages}{6899} (\bibinfo{year}{2001}), \eprint{quant-ph/0105028}.

\bibitem[{\citenamefont{Oppenheim
  et~al.}(2002{\natexlab{b}})\citenamefont{Oppenheim, Horodecki, Horodecki,
  Horodecki, and Horodecki}}]{compl}
\bibinfo{author}{\bibfnamefont{J.}~\bibnamefont{Oppenheim}},
  \bibinfo{author}{\bibfnamefont{K.}~\bibnamefont{Horodecki}},
  \bibinfo{author}{\bibfnamefont{M.}~\bibnamefont{Horodecki}},
  \bibinfo{author}{\bibfnamefont{P.}~\bibnamefont{Horodecki}},
  \bibnamefont{and} \bibinfo{author}{\bibfnamefont{R.}~\bibnamefont{Horodecki}}
  (\bibinfo{year}{2002}{\natexlab{b}}), \eprint{quant-ph/0207025}.

\bibitem[{\citenamefont{Terhal et~al.}()\citenamefont{Terhal, Horodecki,
  Vincenzo, and Leung}}]{IBMHor2001}
\bibinfo{author}{\bibfnamefont{B.~M.} \bibnamefont{Terhal}},
  \bibinfo{author}{\bibfnamefont{M.}~\bibnamefont{Horodecki}},
  \bibinfo{author}{\bibfnamefont{D.~P.~D.} \bibnamefont{Vincenzo}},
  \bibnamefont{and} \bibinfo{author}{\bibfnamefont{D.}~\bibnamefont{Leung}},
  \eprint{quant-ph/0202044}.

\bibitem[{\citenamefont{DiVincenzo et~al.}()\citenamefont{DiVincenzo,
  Horodecki, Leung, Smolin, and Tehral}}]{Locking}
\bibinfo{author}{\bibfnamefont{D.}~\bibnamefont{DiVincenzo}},
  \bibinfo{author}{\bibfnamefont{M.}~\bibnamefont{Horodecki}},
  \bibinfo{author}{\bibfnamefont{D.}~\bibnamefont{Leung}},
  \bibinfo{author}{\bibfnamefont{J.}~\bibnamefont{Smolin}}, \bibnamefont{and}
  \bibinfo{author}{\bibfnamefont{B.}~\bibnamefont{Tehral}},
  \eprint{quant-ph/0303088}.

\bibitem[{\citenamefont{Devetak and Winter}()}]{devwin}
\bibinfo{author}{\bibfnamefont{I.}~\bibnamefont{Devetak}} \bibnamefont{and}
  \bibinfo{author}{\bibfnamefont{A.}~\bibnamefont{Winter}},
  \eprint{quant-ph/0304196}.

\bibitem[{\citenamefont{Devetak}(2003)}]{devetakC}
\bibinfo{author}{\bibfnamefont{I.}~\bibnamefont{Devetak}}
  (\bibinfo{year}{2003}), \eprint{quant-ph/02304196}.

\bibitem[{\citenamefont{A.Wehrl}(1978)}]{W78}
\bibinfo{author}{\bibnamefont{A.Wehrl}}, \bibinfo{journal}{Rev.Mod.Phys.}
  \textbf{\bibinfo{volume}{50}}, \bibinfo{pages}{221} (\bibinfo{year}{1978}).

\bibitem[{\citenamefont{Schumacher and Westmoreland}(2001)}]{SW99}
\bibinfo{author}{\bibfnamefont{B.}~\bibnamefont{Schumacher}} \bibnamefont{and}
  \bibinfo{author}{\bibfnamefont{M.}~\bibnamefont{Westmoreland}},
  \bibinfo{journal}{Phys.Rev. A} \textbf{\bibinfo{volume}{63}},
  \bibinfo{pages}{022308} (\bibinfo{year}{2001}), \eprint{quant-ph/9912122}.

\bibitem[{\citenamefont{King et~al.}(2002)\citenamefont{King, Nathanson, and
  Ruskai}}]{KingNR2001}
\bibinfo{author}{\bibfnamefont{C.}~\bibnamefont{King}},
  \bibinfo{author}{\bibfnamefont{M.}~\bibnamefont{Nathanson}},
  \bibnamefont{and} \bibinfo{author}{\bibfnamefont{M.~B.}
  \bibnamefont{Ruskai}}, \bibinfo{journal}{Phys. Rev. Lett.}
  \textbf{\bibinfo{volume}{88}}, \bibinfo{pages}{057901}
  (\bibinfo{year}{2002}), \eprint{quant-ph/0109079}.

\bibitem[{\citenamefont{Fuchs}(1997)}]{Fuchs97}
\bibinfo{author}{\bibfnamefont{A.}~\bibnamefont{Fuchs}},
  \bibinfo{journal}{Phys. Rev. Lett.} \textbf{\bibinfo{volume}{79}},
  \bibinfo{pages}{1162} (\bibinfo{year}{1997}).

\bibitem[{hug()}]{huge}
\bibinfo{note}{M.Horodecki and P. Horodecki and J.Oppenheim and A.Sen(De) and
  U.Sen and B.Synak. In preparation}.

\end{thebibliography}
\end{document}